\documentclass[10pt,journal,compsoc]{IEEEtran}

\ifCLASSOPTIONcompsoc
  \usepackage[nocompress]{cite}
\else
  \usepackage{cite}
\fi
\ifCLASSINFOpdf
\else
\fi
\usepackage{amsmath}
\usepackage{setspace}
\usepackage{graphicx} 
\usepackage{float} 
\usepackage{subfigure} 
\usepackage{multirow}
\usepackage{diagbox}
\usepackage{color}
\usepackage{enumitem}
\usepackage{cleveref}

\begin{document}
%
\title{Signed Latent Factors for Spamming Activity Detection \> }
%
%
%
%

\author{Yuli Liu
\IEEEcompsocitemizethanks{\IEEEcompsocthanksitem Yuli Liu is with Quan Cheng Laboratory, Jinan, China, as well as the School of Computer Technology and Applications, Qinghai University, Xining, China. \protect\\
E-mail: liuyuli012@gmail.com
}
\thanks{Manuscript received April 19, 2005; revised August 26, 2015.}}

%
%

\markboth{Journal of \LaTeX\ Class Files,~Vol.~14, No.~8, August~2015}%
{Shell \MakeLowercase{\textit{et al.}}: Bare Advanced Demo of IEEEtran.cls for IEEE Computer Society Journals}
%



\IEEEtitleabstractindextext{%
\begin{abstract}
Due to the increasing trend of performing spamming activities (\textit{e.g.}, Web spam, deceptive reviews, fake followers, etc.) on various online platforms to gain undeserved benefits, spam detection has emerged as a hot research issue. Previous attempts to combat spam mainly employ features related to metadata, user behaviors, or relational ties. These studies have made considerable progress in understanding and filtering spamming campaigns. However, this problem remains far from fully solved. Almost all the proposed features focus on a limited number of observed attributes or explainable phenomena, making it difficult for existing methods to achieve further improvement. 
  
  To broaden the vision about solving the spam problem and address long-standing challenges (class imbalance and graph incompleteness) in the spam detection area, we propose a new attempt of utilizing signed latent factors to filter fraudulent activities. The spam-contaminated relational datasets of multiple online applications in this scenario are interpreted by the unified signed network. Two competitive and highly dissimilar algorithms of latent factors mining (LFM) models are designed based on multi-relational likelihoods estimation (LFM-MRLE) and signed pairwise ranking (LFM-SPR), respectively. We then explore how to apply the mined latent factors to spam detection tasks. Experiments on real-world datasets of different kinds of Web applications (social media and Web forum) indicate that LFM models outperform state-of-the-art baselines in detecting spamming activities. By specifically manipulating experimental data, the effectiveness of our methods in dealing with incomplete and imbalanced challenges is validated. 
\end{abstract}

}

\maketitle

\IEEEdisplaynontitleabstractindextext

%
\IEEEpeerreviewmaketitle

\ifCLASSOPTIONcompsoc
\IEEEraisesectionheading{\section{Introduction}\label{sec:introduction}}
\else
\section{Introduction}
\label{sec:introduction}
\fi

%
%
%
%

 

A wide variety of spamming activities have exerted enormous negative impacts on Web platforms. We list some examples to introduce these fraudulent campaigns: (\romannumeral1) \textsl{fake following relationships} on social networking sites. To gain undeserved fans (or excessive influence) for promoting products or services \cite{almaatouq2014twitter, liu2016pay}, many malicious users turn to the underground follower market to buy traded followers, disrupting the fair following mechanism; (\romannumeral2) \textsl{opinion spam} on review websites \cite{jindal2008opinion, mukherjee2013yelp}. If spammers unfairly evaluate targeted (cheating) items, the manipulated sentiment of user-generated contents would likely be instilled to prospective consumers, affecting their final decisions \cite{luca2016reviews}; (\romannumeral3) \textsl{Web spam} refers to activities that use various techniques to make spam Web pages achieve higher-than-deserved rankings in a search engine's result list \cite{gyongyi2004combating, liu2008identifying}. 

\begin{figure}
  \centering
  \includegraphics[width=.85\linewidth]{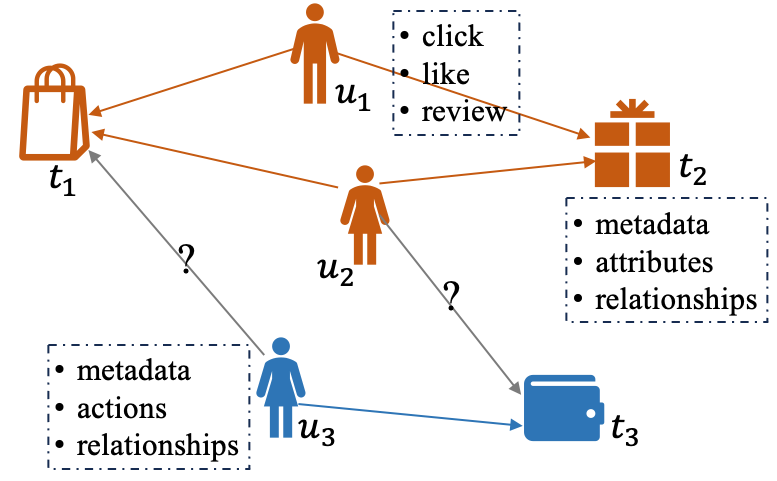}
  \vspace{-0.5mm}
  \caption{ The network diagram shows interactions between users and products. Orange nodes signify fraudulent users and products, while blue nodes denote normal ones. Solid orange lines indicate known fraudulent interactions, and solid blue lines represent legitimate ones. Interactions marked with question marks denote connections with unknown attributes.}
  \vspace{-4mm}
  \label{fig:example}
\end{figure}

Existing anti-spam approaches mainly fall into three categories: (\romannumeral1) \textsl{Behavior Analysis}, leveraging patterns of conducts (\textit{e.g.}, rating patterns of reviews \cite{kumar2018rev2, jindal2010finding} and clicking attributes of search results \cite{liu2008identifying}), behavioral properties (\textit{e.g.}, synchronicity and normality \cite{jiang2014catchsync}), etc. to catch abnormal behavior signals; (\romannumeral2) \textsl{Content Processing}, extracting distinctive characteristics by profile analysis \cite{aggarwal2015they, shah2010evaluating}, semantic representation \cite{sandulescu2015detecting, kim2015deep, ren2016deceptive} and syntactic structure \cite{ fairbanks2018credibility, kennedy2019fact} for filtering spam; (\romannumeral3) \textsl{Relational Structure}, label propagation methods (\textit{e.g.}, bipartite propagation with attribute-related weights \cite{zhang2015catch}, Markov Random Fields with intuitive parameters \cite{rayana2015collective}) and graph properties \cite{noekhah2020opinion, li2019crowdguard} have been fully utilized. 

These works lead us to know that existing detection methods are mainly implemented based on noticeable or explainable clues/ties extracted from multi-source information of nodes and connection relationships in interaction networks as depicted in the example within \Cref{fig:example}. These studies produce a limited number of features for dealing with a specific type of spamming activity. Consequently, ideas of designing detection models are hard to stimulate and performances of combating spam are restricted \cite{liu2020recommending, zhang2015catch}. To overcome the limits of mainstream methods, we intend to combat spamming activities from a fresh perspective, \textit{i.e.}, applying latent factors to spam detection tasks. 
In the recommendation area, using matrix factorization (MF) methods to acquire latent factors for predicting users' preferences has become a common practice \cite{he2017neural}. This is because latent factors represent potential interests (features) of users (items), and expand known information by providing implicit indications. 
However, spam detection tasks pose a more complex challenge: the interaction networks contain both normal (positive) and spam (negative) instances. Consequently, if traditional MF is used in this context, it treats all interactions (whether positive or negative) as equivalent, potentially leading to misclassification. For example, as shown in \Cref{fig:example}, the model might wrongly classify a legitimate interaction between $u_3$ and $t_1$ as spam, given that one-directional latent factors in MF tend to mine similar interaction patterns without distinguishing intent. 
To overcome this, we treat the contaminated relational datasets as signed networks, where each entity has two distinct sets of latent factors, one representing positive relationships and the other representing negative relationships. This perspective, which we term Signed \underline{L}atent \underline{F}actor \underline{M}ining (LFM), enables each latent factor to specialize in modeling either genuine user behavior or malicious activities, thereby enhancing the model's capability to accurately distinguish normal interactions from spam, even within complex, mixed-interaction networks.
In LFM scenario, no explicit feature is required, but implicit signals are mined to filter malicious campaigns. 

We develop two competitive algorithms to mine spam-detection oriented signed latent factors: (\romannumeral1) \textsl{Pointwise Learning} --- \textbf{m}ulti-\textbf{r}elational \textbf{l}ikelihoods \textbf{e}stimation (LFM-MRLE), and (\romannumeral2) \textsl{Pairwise Learning} --- \textbf{s}igned \textbf{p}airwise \textbf{r}anking (LFM-SPR). These two methods (MRLE and SPR) consider \textbf{null} relationships (non-linked node pairs) and \textbf{label unknown} activities (\textit{i.e.}, unlabeled data or labeled instances in test dataset), respectively. Three benefits are brought by designing different LFM models: (\romannumeral1) It validates that this new perspective of LFM possesses considerable potentials and capabilities for combating spamming campaigns (evidences in Section 5.3); (\romannumeral2) We can analyze the influence of null and label unknown information based on experimental results; (\romannumeral3) Combining signed latent factors of MRLE and SPR can boost the final detection performance. 

The other intuitions of applying the LFM model derive from the intention of tackling long-standing challenges in the spam detection area: (\romannumeral1) (\textsl{Annotation Difficulty}) Because spamming campaigns in recent years are usually crowdsourced form \cite{liu2017detecting}, the deliberate camouflage of spammers superficially increases the difficulty of labeling malicious instances. This causes two intractable problems: the lack of training labels \cite{mukherjee2013yelp, rayana2015collective} and class imbalance (\textit{i.e.}, the imbalanced ratio between known spam instances and normal/unlabeled data) \cite{mukherjee2013yelp, fayazi2015uncovering}. In this work, LFM models make full use of different kinds of relationships (\textit{e.g.}, non-linked pairs and label unknown links), and extract expanded potential factors from known information, providing feasibility of tackling these problems; 
(\romannumeral2) (\textsl{Incomplete Graph Structure}) It is unlikely for researchers to crawl and analyze the whole online network and metadata in a website, due to insufficient permissions and computations. In this scenario, Web users or other instances in experimental datasets inevitably lose some relational ties or signals that may be useful for identifying spamming activities \cite{rout2017deceptive}. Furthermore, numerous collected users may only be related to a few numbers of items/targets in the experimental graph, generating a highly sparse relational matrix, which further aggravates the degree of graph incompleteness. Another motivation to introduce the thought of latent factors to the anti-spam area is evoked, since it is well known that the latent factors of MF are mainly used to learn incomplete or potential interactions \cite{krishna2018simple, zhang2018sine}. 

In recent years, deep learning technologies, particularly Graph Neural Networks (GNNs), have been extensively applied in fraud detection tasks. GNN methods enhance the expressivity of latent representations by propagating and aggregating information across nodes, thereby capturing structural \cite{wu2023splitgnn, danilchenko2022opinion, liu2023improving} and feature-based properties \cite{cai2024detecting, 10574870, 10458116} of networks. While these latent representations may not be intuitively explainable, they are derived from and depend on the underlying features of the network, such as node connectivity, edge weights, and node attributes. 
Essentially, these cutting-edge graph-based models methods are based on neural network structures that improve representations. They integrate relational structures and graph/node features into a cohesive neural network model. In contrast, our LFM methods directly optimize signed latent factors, fundamentally acting as an optimization criterion tailored for spam detection tasks. Consequently, we can employ neural network architectures to serve as a front-end input to LFM, which enhances the initial representations that our LFM approaches subsequently refine and optimize. This implies that the LFM optimization approaches can improve existing deep neural network-based models for spam detection. The applicability of this approach has been validated through experiments.

It has been reported that designing a specific detection model to deal with a certain kind of spamming activity makes researchers exhausted to tackle the continuously emerging spamming phenomena \cite{liu2020recommending}. Besides, feature-based detection models suffer from limited robustness and capability, since even the same type of spamming campaign frequently changes its cheating strategies \cite{liu2016pay, zhang2015catch}. To cope with these predicaments, relational datasets of different kinds of fraudulent phenomena in this work are interpreted by the unified signed network, and then LFM models that involve no specific features can be generalized to detect disparate spamming activities.

We summarize the contributions of this work as follows: 
\begin{itemize} 
\item We uniformly use the signed network to represent various spam-contaminated datasets and introduce a new perspective (\textit{i.e.}, LFM) to address the spam detection problem.
\item Two competitive and highly dissimilar algorithms, SPR and MRLE, are proposed to mine spam-detection oriented signed latent factors, tackling the long-standing challenges. 
\item To explore the capability of signed latent factors for filtering spamming activities, five types of application strategies are proposed and evaluated. 
\item Our LFM approach can be viewed as an optimization criterion specifically designed for detecting fraudulent activities, capable of enhancing existing neural network-based spam detection architectures. Experimental results have demonstrated the applicability of LFM, showcasing its ability to refine and optimize detection mechanisms effectively. We will release the code upon acceptance.
\end{itemize}

\section{SIGNED LATENT FACTOR MINING}
To increase the generality and usability of signed latent factors for spam detection (\textit{i.e.}, detecting multiple kinds of spamming activities), we use the unified relational structure to represent disparate contaminated networks. Due to the interactivity between users and other targets of different online platforms \cite{liu2020recommending}, spam-contaminated networks can be uniformly treated as the signed user-target relational structure. In this scenario, targets can represent products, Web pages, videos, etc. that users interact with on Web applications. 

Existing detection methods mainly focus on filtering three types of spamming objectives: spammers \cite{wang2011review}, malicious activities \cite{li2019crowdguard} or cheating targets \cite{liu2016pay}. In this work, we aim to filter the second type, because detecting fraudulent activities facilitates timely eliminating and alleviating negative influences of malicious phenomena \cite{kim2015deep, liu2017detecting}. In the signed user-target structure, edges/entries are mainly constructed by activities, \textit{e.g.}, the user-review-product structure on review websites, user-click-page relationships of search engines, and user-like-post networks on social media platforms. Therefore, detecting spamming activities in this paper means identifying abnormal edges (entries) in signed networks (matrix). Besides, LFM models only rely on ties/relations between users and targets to mine signed latent factors, without using any crafted features. It means that spammers can not easily avoid LFM models’ detection by changing disguise strategies, because inherent connections between spammers and malicious targets are hard to conceal \cite{liu2016pay,gong2014sybilbelief}. 

\subsection{Application Strategies}

\begin{table}[tp]
\centering
  \fontsize{7}{9}\selectfont
  \caption{Statistics of the datasets.}
  \setlength{\tabcolsep}{0.1mm}{
  \label{tab:performance_comparison}
    \begin{tabular}{ccc}
    \hline
Notation&Description&Dimensionality\\
    \hline
    Average ($Avg$) & $f(u,t)=\frac{1}{2}(W_u+H_t)$ & $d^+ + d^-$\\
    Concatenate ($Con$) & $f(u,t)=(W_u, H_t)$ &  $2*(d^+ + d^-)$ \\  
    Subtraction ($Sub$) & $f(u,t)=(W_u^- - W_u^+, H_t^- - H_t^+)$ & \ $d^+ + d^+$ \\
    Spam Inner product ($IP^-$)  & $f(u,t)=W_u^- H_t^{-T}$ & $\ \ 1$\\
    Normal Inner product ($IP^+$)  & $f(u,t)=W_u^+ H_t^{+T}$ & $\ 1$\\ \hline
    \end{tabular}}
\end{table}

\begin{figure*}
  \centering
  \includegraphics[width=0.95\linewidth]{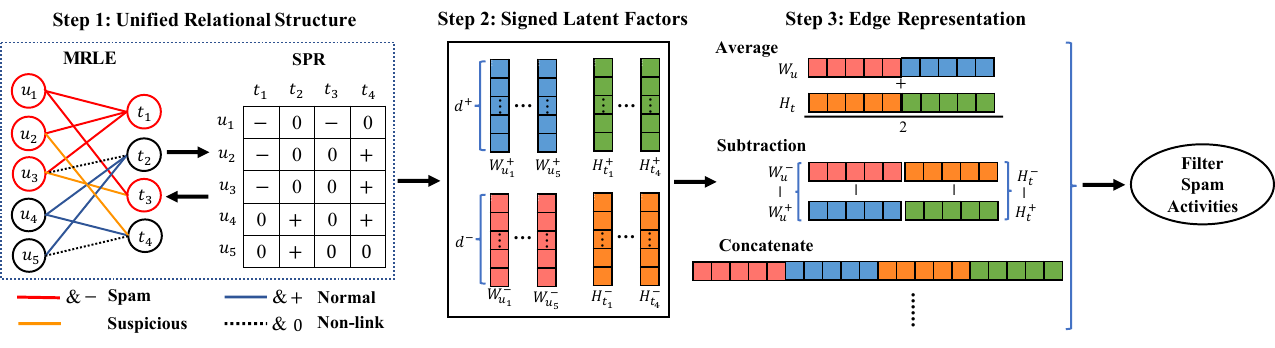}
  \caption{The application procedure of signed latent factors for spamming activity detection.}
  \label{fig:application}
  \vspace{-2mm}
\end{figure*}

In the recommendation area, using low-dimensional latent factors obtained from matrix factorization to represent potential preferences (features) of users (items) is a popular method. As we know, the recommender system aims to predict satisfying items for users, regardless of the moral label of a user/item (\textit{i.e.}, without considering whether a user/item is benign or malicious). Therefore, the latent factors mined for recommendation have no normal or spamming sign (\textit{i.e.}, each user/item is usually represented by one vector). However, in the area of spam detection, all instances are divided into two classes, normal (positive) and spamming (negative). Consequently, to capture two classes of latent factors corresponding to the normal and spamming indications of each instance, all users and targets in the unified structure are characterized by two vectors, \textit{i.e.}, positive factors $W_u^+$ and $H_t^+$ for all users and targets (dimensions of positive latent factors are $d^+$), and negative forms $W_u^-$ and $H_t^-$ of users and targets (dimensions are both $d^-$). 

As mentioned above, we focus on detecting spamming activities, \textit{i.e.}, edges (entries) of a graph (matrix). Referring to previous work \cite{xu2019link, su2018detecting}, the class of an activity on Web applications is determined by the linked user and target. Therefore, after obtaining the signed latent factors of nodes (entities), we need to explore how to comprehensively apply them to filtering malicious activities. In this work, five kinds of application modes are proposed to utilize the signed latent factors of linked nodes (user and target) to represent a link (activity), which are listed in Table 1: (\romannumeral1) Average ($Ave$), calculating the average results after adding $W_u$ and $H_t$, where $W_u$ and $H_t$ denote the concatenation of user latent factors ($W_u^+$ and $W_u^-$) and target latent factors ($H_t^+$ and $H_t^-$), respectively; (\romannumeral2) Concatenate ($Con$), all kinds of latent factors are directly and integrally concatenated. This type of operator produces the combined factor vectors with the widest dimensionality --- $2*(d^+ + d^-)$; (\romannumeral3) Subtraction ($Sub$), using the spamming indications of user and target ($W_u^-$ and $H_t^-$) to subtract the normal indications ($W_u^+$ and $H_t^+$), with the intention of avoiding misjudging normal activities as spam ones. We need to note that the dimensions of positive and negative factors should be the same when we use $Ave$ and $Sub$ operators. 

The above-mentioned three application modes utilize the combined factors as features and feed them into a classification model to filter spamming activities. The other two operators based on calculating the inner product are inspired by the common practice of MF recommendation. The fourth operator $IP^-$ denotes the inner product of $W_u^-$ and $H_t^-$, representing the possibility of the edge ($u,t$) being deceptive (negative). Similarly, the result of $IP^+$ denotes the normal possibility of the corresponding edge. The negative and positive inner product results of all the activities in datasets can be ranked in descending order, and we hold that spamming activities will rank high (low) in the $IP^-$ ($IP^+$) result list.

\Cref{fig:application} demonstrates the 3-step procedure of applying signed latent factors to detect spamming activities. Two forms of unified user-target relational structures, \textit{i.e.}, signed social graph and signed matrix for MRLE model and SPR model, respectively, are constructed to capture signed signals. In the bipartite graph (adjacency matrix), red solid lines (-) and blue solid lines (+) represent known fraudulent and normal behaviors, respectively. Orange solid lines indicate suspicious relationships, as fraudulent users might interact with legitimate items to disguise their activities. Dashed lines and zeros denote the absence of a connection. 
In the pointwise method (MRLE), we construct signed networks based on the complete set of interaction relationships. Using labeled interactions, we define three distinct types of edges within these networks (detailed in Section 3), which enable us to calculate the likelihood of interactions, thereby deriving optimized signed latent factors. For the signed pairwise ranking (SPR) model, we create a signed sparse matrix representation and construct sufficient comparison pairs from positive and negative perspectives, respectively (detailed in Section 4). This process facilitates ranking optimization, ultimately yielding refined signed latent factors.
The third step in \Cref{fig:application} presents some toy examples to illustrate the proposed operators of edge (activity) representations for filtering spamming actions using mined latent factors of connected nodes (user and target). These activity representations are then classified using existing classification techniques to distinguish potential spamming activities.

\section{LFM-MRLE MODEL}
We propose the first LFM model --- multi-relational likelihoods estimation (MRLE) on the basis of defining three types of naturally existing relations in spam-related networks inspired by the link prediction work \cite{xu2019link}: (\romannumeral1) normal/positive relations ($nor$), \textit{i.e.}, benign activities; (\romannumeral2) spam/negative ($sp$) ties; (\romannumeral3) non-linked ($non$) node pairs, \textit{i.e.}, there is no interaction between users and targets. In spam detection area, the null relationships are often ignored by existing detection models. 

After defining three types of spam-related relationships, the next key step is to quantify them through three types of likelihood formulations based on analyzing the interplay between spam and normal indications of each node pair. To quantify the indicative meanings of likelihoods and obtain nonlinear latent factors, we utilize the activation function $f_a(x) = \frac{p_0exp(x)}{1+p_0(exp(x)-1)}$, which is similar to the common sigmoid function in neural networks. Given a node pair $(u,t)$, we use $F_{ut}^- = f_a(W^-_u H^-_t$) to denote the calculated spamming signal, where $W^-_u H^-_t$ is the inner product of spam latent representations of user $u$ and target $t$ in user spam latent factor matrix $W^-$ and target spam latent factor matrix $H^-$. Similarly, we can also capture the quantified normal signal, using $F_{ut}^+ = f_a(W^+_u H^+_t$). 

We now describe how to formulate three types of relations and explain corresponding indicative meanings: (\textbf{\romannumeral1}) (\textbf{nor}mal link) If an edge $(u,t)$ (\textit{i.e.}, activity performed by $u$ to $t$) is normal, the user $u$ and target $t$ are connected by a benign tie, and they are both inclined to be normal. It means that the value of normal indication $F_{ut}^+$ should be large and $F_{ut}^-$ is assumed to be small, so we combine them with the goal of making the score of $F_{ut}^+(1-F_{ut}^-)$ large; (\textbf{\romannumeral2}) (\textbf{sp}am link) Oppositely, in this case, the $u$ and $t$ are associated with fraudulent activity, \textit{i.e.}, the possibility for corresponding link being spam are high. So we use $(1-F_{ut}^+)F_{ut}^-$ to represent the spam likelihood; (\textbf{\romannumeral3}) (\textbf{non}-linked pair) Different from most previous works, the unconnected relationship is also considered in this work. We tend to believe that null relationships capture the possible discrepancy between corresponding node pairs, and signify that there is neither the positive nor negative relation between $u$ and $t$, which can be represented by $(1-F_{ut}^+)(1-F_{ut}^-)$. 
Besides, three types of edge sets are represented by $E^{nor}$, $E^{sp}$ and $E^{non}$, respectively. 

\subsection{Estimating Multi-Relational Likelihoods}
We integrate multi-relational likelihoods using the maximum likelihood formulation, and employ negative log-likelihood objective function to estimate signed latent factors, defined as follows:

\begin{equation}
\begin{aligned}
  L = & - \sum\nolimits_{(u,t)\in E^{nor}}\log(F_{ut}^+(1-F_{ut}^-)) \\ &- \sum\nolimits_{(u,t)\in E^{sp}}\log ((1-F_{ut}^+)F_{ut}^-) \\ &- \sum\nolimits_{(u,t)\in E^{non}}\log((1-F_{ut}^+) (1-F_{ut}^-)).
\end{aligned}
\vspace{1mm}
\end{equation}
This objective function is composed of three components, corresponding to three sets of relationships. 

To calculate signed latent factors of each node in the unified signed network, the components based on edge form need to be transformed into the node-based format. As there are two types of entities in relational data: user and target, the integrated formulation is then divided into two parts, \textit{i.e.}, user and target forms. The user-perspective one is formulated as: 

\begin{equation}
\begin{aligned}
   L_{(u)} = &- \sum\nolimits_{t\in T^{nor}(u)}\log(F_{ut}^+(1-F_{ut}^-)) \\ & - \sum\nolimits_{t\in T^{sp}(u)}\log((1-F_{ut}^+)F_{ut}^-)  \\ & - \sum\nolimits_{t\in T^{non}(u)}\log((1-F_{ut}^+)(1-F_{ut}^-)),
\end{aligned}
\vspace{1mm}
\end{equation}
where $T$ denotes the set of targets, and $T^{nor}(u)$, $T^{sp}(u)$, and $T^{non}(u)$ represent the sets of targets that respectively receive normal, spam, and null connections from user $u$, \textit{i.e.}, three types of relations. We can see that a user's likelihood is calculated based on all targets in the signed network that have different relationships with it. In real-world networks, the majority of targets have no link to a specific user. If all of them are considered when we estimate the user's $non$ likelihood, a huge amount of computational resources will be consumed. Therefore, in the process of calculating the null relational likelihood, we only sample $n$ targets that are not associated with the user. Through experiments of fixing other parameters but vary $n$, the effect of considering null relationships in detecting spamming activities is investigated (shown in Section 5.3). Similarly, the other objective function focusing on targets can be formulated by replacing $u$ to $t$, or conversely. To save space, the corresponding formulation is not presented. 

To minimize the objective functions, the coordinate descent method \cite{lin2007projected, xu2019link} is used, where signed latent vectors of an instance are updated while fixing all the other vectors in each iteration. The minimization task then becomes a convex optimization problem. The derivatives of a user's signed latent factors are:

\begin{equation}
\begin{aligned}
  \frac{\partial L_{(u)}}{\partial W_u^+}  =  & - \sum\nolimits_{t\in T^{nor}(u)} ((1-F_{ut}^+)H_t^+) \\& +  \sum\nolimits_{t\in T^{sp}(u)\cup T^{non}(u)}F_{ut}^+H^+_t,
\end{aligned}
\end{equation}

\begin{equation}
\begin{aligned}
  \frac{\partial L_{(u)}}{\partial W_u^-}  = & - \sum\nolimits_{t\in T^{sp}(u)}((1-F_{ut}^-)H_t^-)  \\& +  \sum\nolimits_{t\in T^{nor}(u)\cup T^{non}(u)}F_{ut}^-H^-_t.
\end{aligned}
\vspace{1mm}
\end{equation}
The other two derivatives, \textit{i.e.}, normal and spam target latent factors ($H^+$, $H^-$), can be easily deduced based on the equations above. 

Additionally, to validate the convergence of our MRLE model, we implement the L-BFGS-B algorithm \cite{zhu1997algorithm}. By performing experiments on a real-world dataset used by Liu \textsl{et al.} \cite{liu2016pay} described in Section 5.1, we find that the optimization algorithm successfully achieves convergence after only 72 times of iterations.

In the calculating process, we find that LFM-MRLE estimates the single predictor (\textit{i.e.}, individual latent vector as is shown in Equation 3 and 4), without calculating the difference between two predictors. Besides, although the MRLE method considers non-linked relationships in the learning process, the \textbf{label unknown} activities (unlabeled instances and labeled data used for testing in signed networks) are omitted. In the spam detection area, all existing connections (including label unknown ones) are useful, because they keep the integrity of networks and facilitate extracting distinctive structure properties for detecting spam \cite{li2017bimodal, jiang2014catchsync}.  
In view of these two problems, we try to take comparative ranking information into account (\textit{i.e.}, to estimate the difference of two predictors) on the integral network. From this side, the commonly used objective function --- pairwise loss in BPR method  \cite{rendle2009bpr} comes to mind, which is still a highly competitive approach in item recommendation \cite{he2017neural, liu2024learning}. 

\section{LFM-SPR MODEL}
We now start to introduce the preliminaries of LFM-SPR model based on signed pairwise relations. The signed user-target matrix $Y$: $U\times T$, where $U$ and $T$ represent user and target sets, is defined according to the signed network as,
\begin{equation}
y_{ut} =\left\{
\begin{aligned}
+,      &\ \ \ \ \ \ \text{ if user } u \text{ perform a normal activity to target } t; \\
-,      &\ \ \ \ \ \ \text{ if user } u \text{ perform a spam activity to target } t;\\
?,      &\ \ \ \ \ \ \text{ if label of interaction ( } u, t \text{ ) is unknown }; \\
0,      &\ \ \ \ \ \ \text{ no link between } u \text{ and } t.
\end{aligned}
\right.
\end{equation}  
Here the matrix is quite different from the common interpretation of recommendation works called 0/1 scheme \cite{rendle2009learning, liu2024pay} that uses value $1$ to indicate the observed interaction and $0$ to denote the unobserved one. The reason of defining four types of entries mainly lies in: (\romannumeral1) The spam-contaminated matrix is signed, with normal ($+$) and spam ($-$) entries; (\romannumeral2) To consider all existing connections into signed latent factors mining for combating spam activities, the \textbf{label unknown} links ($?$) are also used to provide additional clues for comparing signed ranking pairs. 

\begin{figure}
  \centering
  \includegraphics[width=1.\linewidth]{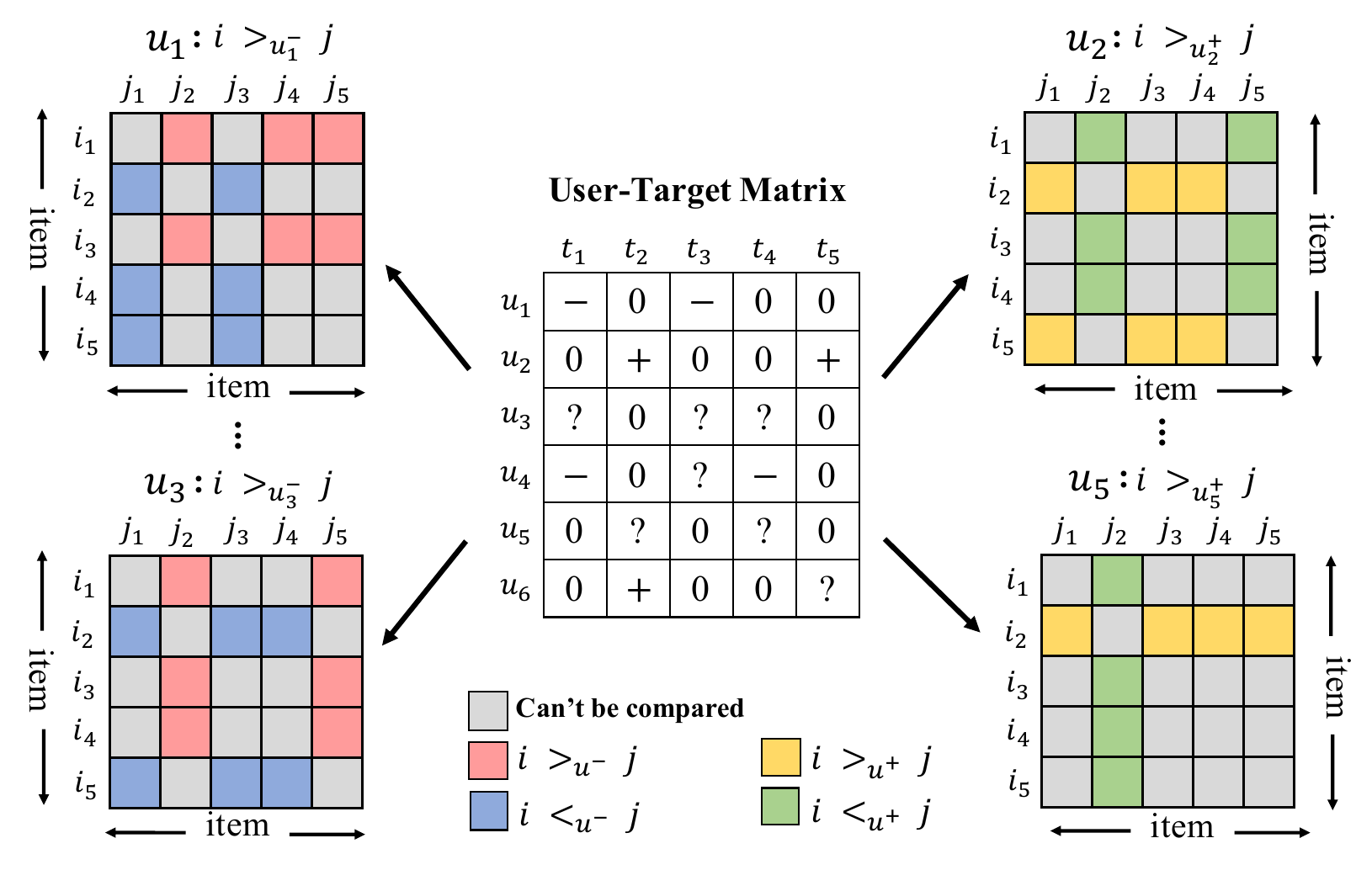}
  \caption{Motivating Example of Signed Pairwise Ranking}
  \label{fig:spr-example}
\end{figure}

In order to clarify the SPR model, we use \Cref{fig:spr-example}  as a motivating example. The symbols $+$ and $-$ denote known normal and spamming activities, \textit{i.e.}, labeled edges used as training data. A value of $0$ indicates the null relationship. The entries with the symbol of $?$ are composed of two types of instances: (\romannumeral1) unlabeled activities, and (\romannumeral2) labeled activities used as test data, \textit{i.e.}, entries with $?$ are unknown for SPR learning. Usually, in recommendation tasks, the observed feedbacks divided into the test dataset are directly treated as unobserved ones in the learning process, which will be used to evaluate performances of the recommender system \cite{he2017neural, liu2024probabilistic}. However, it is necessary to set a new type of entries (\textit{i.e.}, $?$) to represent the unique instances in spam detection tasks: (\romannumeral1) There are unlabeled datasets in many works of filtering spam due to the difficulty of labeling instances (\textit{i.e.}, Annotation Difficulty), which are unable to be treated as positive nor negative ones; (\romannumeral2) If we use the common practice of recommendation tasks to directly handle the unknown connections as unobserved feedbacks ($0$), these existing but label unknown links will be blended into null relations. It means that an important type of relational ties that plays a distinctive role in detecting spam \cite{gong2014sybilbelief} will be ignored. 

In real-world spamming activity detection experiments, to fairly evaluate the model, randomly dividing training and test datasets is often used. In this paper, we aim to detect spamming activities rather than malicious users or targets, so all or part of the labeled activities of some users may be used as test data after dividing the dataset, just like $u_3$ (all entries are unknown) and $u_4$ (one entry with $?$) in the matrix of \Cref{fig:application}. However, to train and test the MF model for a personalized recommendation, it is quite common for researchers to randomly select a part of each user's feedbacks as test data \cite{rendle2009bpr, he2017neural, liu2024probabilistic}, \textit{i.e.}, each user contains observed feedbacks. This makes predicting and testing recommendation results possible. As for spamming activity detection tasks, each user has unconcealed activities in the training process is unfair for evaluation. Because spammers (benign users) mainly perform spam (normal) behaviors \cite{lu2013simultaneously, liu2016pay}, the label of a user's known activities is a strong clue for a detection model to give the same class for the user's other unknown activities as the label of known ones.

The recommendation problem usually can be abstracted as learning $\hat{y_{ui}}=f(u,i|\Theta)$, where $\hat{y_{ui}}$ indicates the predicted score of an interaction between user $u$ and item $i$ \cite{he2017neural}, and $\Theta$ represents the parameter vector of an arbitrary models class (\textit{e.g.}, matrix factorization). And two types of objective function are most widely used to estimate $\Theta$ in existing literature --- pointwise loss \cite{hu2008collaborative, he2017neural} and pairwise loss \cite{rendle2009bpr}. 
To formulate a pairwise loss function for an item recommendation, the first step is to construct pairwise relations which are usually based on specific user's preferences, \textit{i.e.}, a user prefers observed feedbacks over unobserved ones \cite{rendle2009bpr}. Therefore, the matrix factorization of BPR only needs to learn one type of latent factors according to user preferences. As for the contaminated matrix, however, to mine signed latent factors, pairwise relations are built from two perspectives: normal ($+$) level and spam ($-$) level. Through this way, we can not only know spamming pairwise rankings to learn negative latent factors but also can learn positive ones based on normal ranking relations. Consequently, each user/target is represented by two low-dimensional factors. 

Overall, the signed pairwise learning method for spam detection is quite different from BPR for the recommendation, not only in the objective but also in data processing and ranking pair comparison (summarised in the Related Work section). The design of SPR shows that the thought of mining spam detection-oriented signed latent factors is not restricted to a certain direction. Two implementations of LFM signify that more possibilities and insights for using latent factors to solve the spam problem are provided.

\subsection{Constructing Ranking Pairs}

On the left and right sides of \Cref{fig:spr-example}, two levels of ranking relations are extracted for specific users, \textit{i.e.}, spam ($>_u^-$) and normal ($>_u^+$) ranking relations. From spam level, if a target $t$ has received a spam ($-$) action from user $u$, \textit{i.e.}, a negative ($u$, $t$) entry in signed matrix, we can assume that user $u$ brings more spamming indications to target $t$ than other targets that have normal or null interactions with $u$. For example, on the top left of \Cref{fig:spr-example}, user $u_1$ performs spamming activities to $t_1$ and $t_3$, but there is no feedback on $t_2$, $t_4$ and $t_5$. So we tend to believe that targets $t_1$ and $t_3$ are more likely to be malicious entities compared to other targets from the perspective of user $u_1$. That is, based on signed feedbacks of $u_1$, six spam ranking relations can be found: $t_1 >_{u_1^-} t_2$, $t_1 >_{u_1^-}t_4$, $t_1 >_{u_1^-}t_5$, and $t_3$ also ranks higher than $t_2$, $t_4$ and $t_5$ from the perspective of $u_1$. 

On the other hand, if a user contains label unknown activities ($?$), it will cause the clues that facilitate obtaining ranking pairs to be unclear. The corresponding user is named \textbf{ranking-unclear} user. To tackle the clue unclear problem, we attempt to seek complementary evidences from existing interactions of related targets that are linked to the \textbf{ranking-unclear} user, \textit{i.e.}, comparing ranking relations based on the interaction distribution of corresponding target columns. Taking $u_3$ for example, we can see that all of its interactions ($u_3$ row) are unknown ($?$) for learning process. In the motivating matrix, target $t_1$ receives a class unknown ($?$) activity from $u_3$, but there are the other two activities from $u_1$ and $u_4$ (in $t_1$ column), \textit{i.e.}, ($u_1$, $t_1$) and ($u_4$, $t_1$), which are known spamming links ($-$). Based on this evidence, we can say that $t_1$ is suspicious due to its involvement in cheating activities. This means that the label unknown feedback ($u_3$, $t_1$) of ranking-unclear user $u_3$ has higher possibility of being malicious than $u_3$'s non-linked relationships ($u_3$, $t_2$) and ($u_3$, $t_5$), \textit{i.e.}, $t_1 >_{u_3^-} t_2$, $t_1 >_{u_3^-}t_5$. However, from the perspective of $u_3$, we cannot deduce the ranking relations between $t_1$, $t_3$ and $t_4$ (targets that receive $?$ interactions from $u_3$), since the columns of $t_3$ and $t_4$ are also related to spamming interactions, \textit{i.e.}, ($u_1$, $t_3$) and ($u_4$, $t_4$). The spam ranking relations of $u_3$ is shown on the bottom left of \Cref{fig:spr-example}.

In summary, the process of acquiring signed ranking relations can be divided into two parts: (\romannumeral1) analyzing class known entries ($+$ or $-$), and (\romannumeral2) in face of unknown interactions ($?$). Complementary evidence from the linked targets should be taken into account when dealing with the label unknown case of a specific user. Besides, it is important to mention that to cheat detection models, spammers occasionally interact with normal targets to disguise themselves (\textit{i.e.}, \textbf{disguised-activities}). To alleviate the perturbations of this disguise strategy, we set an identification threshold $\xi$ \cite{liu2020recommending}. When we assign $\xi = 1$, for example, the entry ($u_3$, $t_3$) will rank lower than ($u_3$, $t_1$) from the perspective of $u_3$, because there is only one spamming interaction in $t_3$ column, \textit{i.e.}, the number of related spamming links ($-$) of $t_3$ is not bigger than $\xi$. It means that the evidence of $t_3$ being fraudulent is not enough. However, $t_1$ receives two spam feedbacks ($>1$), showing severe fraudulence. Therefore, in the process of obtaining ranking pairs, we can regard the label unknown entries (\textit{e.g.}, ($u_3$, $t_1$)) that are related to fraudulent-enough targets (\textit{e.g.}, $t_1$) as \textbf{auxiliary-higher-ranking} interactions. 

Similarly, the pairwise ranking relations from the normal level can also be constructed (shown on the right side of \Cref{fig:spr-example}). 

\subsection{Optimizing Signed Pairwise Ranking}
The widely used pairwise ranking method BPR \cite{rendle2009bpr} uses parameter $\Theta$ to determine personalized rankings, and applies Bayesian formulation to maximize the following posterior probability,
\begin{equation}
\begin{aligned}
  p(\Theta|i>_uj)) \propto p(i>_uj|\Theta)p(\Theta).
\end{aligned}
\end{equation}
Here, $p(i >_u j|\Theta)$ denotes the likelihood function, which captures the individual probability that user $u$ prefers target $i$ over target $j$, \textit{i.e.}, $i$ represents the higher-ranking target and $j$ is the lower one from the view of $u$. And it is defined as: $p(i>_uj|\Theta):=\sigma(\hat{x}_{uij}(\Theta))$, where $\sigma$ denotes the common logistic sigmoid: $\sigma(x)=\frac{1}{1+e^{-x}}$. Further, the $\hat{x}_{uij}(\Theta)$ captures the relationship between target $i$ and target $j$ based on the preferences of user $u$. It can be decomposed into: $\hat{x}_{uij}=\hat{x}_{ui}-\hat{x}_{uj}$. In the pairwise loss paradigm, researchers try to predict the difference between two predictions $\hat{x}_{ui}$ and $\hat{x}_{uj}$, rather than a single predictor $\hat{x}_{ui}$ or $\hat{x}_{uj}$. 

To mine the SPR signed latent factors, signed trainable parameter vectors $\Theta^+$ and $\Theta^-$ are defined, corresponding to two types of ranking relations: $\hat{x}_{uij}^+$ and $\hat{x}_{uij}^-$, \textit{i.e.}, pairwise ranking relations from normal and spam levels. To sum it up, the integrated form of posterior probability optimization used in this paper can be formulated as:
\begin{equation}
\begin{aligned}
&\ln \prod_{(u, i, j) \in Y_S^{+}} p\left(\Theta^{+} \mid \hat{x}_{u i j}^{+}\right) \prod_{(u, i, j) \in Y_S^{-}}\left(\Theta^{-} \mid \hat{x}_{u i j}^{-}\right) \\
&=\sum_{(u, i, j) \in Y_S^{+}} \ln \sigma\left(\hat{x}_{u i j}^{+}\right)+\sum_{(u, i, j) \in Y_S^{-}} \ln \sigma\left(\hat{x}_{u i j}^{-}\right) \\
& \ \ \ \ \ \ \ \ \ +\ln \left(\Theta^{+}\right)+\ln \left(\Theta^{-}\right) \\
&=\sum_{(u, i, j) \in Y_S^{+}} \ln \sigma\left(\hat{x}_{u i j}^{+}\right)+\sum_{(u, i, j) \in Y_S^{-}} \ln \sigma\left(\hat{x}_{u i j}^{-}\right) \\
& \ \ \ \ \ \ \ \ \ -\lambda_{\Theta^{+}}\left\|\Theta^{+}\right\|^2-\lambda_{\Theta^{-}}\left\|\Theta^{-}\right\|^2
\end{aligned}
\end{equation}
where the triple $(u,i,j)\in{Y_S^+}$ means the normal ranking pair $\hat{x}^+_{ui}-\hat{x}^+_{uj}$ in the training dataset of signed matrix $Y$. To complete the Bayesian modeling approach, a general prior density $p(\Theta)$ with zero mean and variance matrix: $\sum(\Theta) = \lambda_\Theta I$ is introduced \cite{rendle2009bpr}. In the practical calculation, $\lambda_\Theta$ can be seen as the model specific regularization parameter. To mine signed latent factor matrices, two specific regularization parameters are defined $\lambda_{\Theta^+}$ and $\lambda_{\Theta^-}$. After calculating the logarithm of Equation 7, the parameters $\Theta^+$ and $\Theta^-$ are separated into two accumulation parts. It means that we can separately and flexibly calculate the positive and negative latent factors. By flipping the signs of all components in Equation 7, the maximum optimization is transformed into the commonly used minimizing task. 

The LFM-SPR model aims to mine positive and negative latent factors for each user and target. In the optimization equation, the user and target are incorporated into pairwise relations, so we first need to decompose the pairwise estimator $\hat{x}_{uij}^+$ into: $W_u^+ H_i^{+T} - W_u^+ H_j^{+T}$ where $W_u^+$, $H_i^+$ and $H_j^+$ denote the normal latent factors of user $u$, target $i$ and target $j$, respectively. Because the calculation of optimizing signed latent factors can be separated, we only present the formulation of normal part. To learn positive signed latent factors from spam-contaminated matrix, the gradient for three parameters $W_u^+$, $H_i^+$ and $H_j^+$ are needed to derive, using the stochastic gradient descent algorithm:
\begin{equation}
\begin{aligned}
  \frac{\partial}{\partial W_u^+}  = \frac{1}{1+e^{W_u^+ H_i^{+T} - W_u^+ H_j^{+T}}} \cdot (H_j^+ - H_i^+) + \lambda_{\Theta^+}W_u^+
\end{aligned}
\end{equation}

\begin{equation}
\begin{aligned}
  \frac{\partial}{\partial H_i^+}  = \frac{1}{1+e^{W_u^+ H_i^{+T} - W_u^+ H_j^{+T}}} \cdot -W_u^+ + \lambda_{\Theta^+}H_i^+
\end{aligned}
\end{equation}

\begin{equation}
\begin{aligned}
   \frac{\partial}{\partial H_j^+}  = \frac{1}{1+e^{W_u^+ H_i^{+T} - W_u^+ H_j^{+T}}} \cdot W_u^+ + \lambda_{\Theta^+}H_j^+
\end{aligned}
\end{equation}

Finally, an update for these three parameters can be performed in each iteration: 
\begin{equation}
\begin{aligned}
  \Theta^+ \Leftarrow  \Theta^+ - \alpha \frac{\partial}{\partial \Theta^+}
\end{aligned}
\end{equation}
where $\alpha$ denotes the learning rate. The spam latent factors can also be calculated via the similar approach. 

Up to this point, two LFM models (MRLE and SPR) have been introduced. We can see that both models aim to capture signed latent representations of each instance, but the basic thoughts are quite different. By utilizing multi-relational likelihoods and signed pairwise rankings, we try to potentially represent each user and target as comprehensively as possible. 

\section{EXPERIMENTS}

\begin{table*}[tp]
  \centering
  \fontsize{8.5}{11}\selectfont
  \caption{AUC performances of compared methods in two datasets with varying size (\%) of labeled data ($n = 20$, $d = 30$, $\xi = 2$).}
  \setlength{\tabcolsep}{2.6mm}{
    \begin{tabular}{cccccc|ccccc}
    \hline
    \multirow{2}{*}{Methods}&\multicolumn{5}{c|}{\textbf{\texttt{Social}}} & \multicolumn{5}{c}{\textbf{\texttt{Forum}}} \cr
    \cline{2-11}& \ \ 20\%\ \ \ & \ \ \ 25\%\ \ \ & \ \ \ 30\%\ \ \ & \ \ \ \ \ 40\%\ \ \ & \ \ \ 50\%\ \ \ & \ \ \ 20\%\ \ \ & \ \ \ \ 25\%\ \ \ & \ \ \ 30\%\ \ \ & \ \ \ 40\%\ \ \ & \ \ \ 50\%\ \ \  \cr\hline
    MRLE$_{Avg}$&0.708&0.733&0.765&0.770&0.792& 0.639&0.665&0.690&0.709&0.716\cr
    MRLE$_{Con}$&0.809&0.832&0.857&0.870&0.876& 0.718&0.733&0.755&0.767&0.779\cr
    MRLE$_{Sub}$&0.716&0.758&0.778&0.784&0.803& 0.640&0.663&0.695&0.715&0.721\cr
    MRLE$_{IP^-}$&0.704&0.715&0.741&0.757&0.763& 0.639&0.642&0.681&0.695&0.709\cr
    MRLE$_{IP^+}$&0.651&0.664&0.685&0.688&0.696& 0.584&0.595&0.617&0.629&0.635\cr\hline
    
    MRLE$-non$&0.734&0.756&0.774&0.792&0.813& 0.661&0.674&0.692&0.706&0.728\cr
    MRLE$+au$&0.812&\textbf{0.847}&\textbf{0.872}&\textbf{0.879}&\textbf{0.886}& 0.725&0.741&\textbf{0.764}&\textbf{0.775}&\textbf{0.781}\cr\hline

    SPR$_{Avg}$&0.710&0.742&0.750&0.762&0.764& 0.656&0.674&0.693&0.707&0.712\cr

    SPR$_{Con}$&\textbf{0.813}&0.842&0.859&0.864&0.871& \textbf{0.732}&\textbf{0.745}&0.752&0.764&0.771\cr
    SPR$_{Sub}$&0.725&0.754&0.771&0.783&0.794& 0.660&0.684&0.708&0.714&0.726\cr
    SPR$_{IP^-}$&0.709&0.720&0.731&0.742&0.757& 0.642&0.659&0.671&0.683&0.691\cr
    SPR$_{IP^+}$&0.642&0.655&0.673&0.689&0.693& 0.562&0.576&0.584&0.607&0.615\cr\hline

    Graph$_{HITS}$&0.690&0.745&0.762&0.775&0.782& 0.620&0.646&0.671&0.680&0.687\cr
    Graph$_{Bi}$&0.698&0.750&0.774&0.783&0.791& 0.637&0.659&0.679&0.685&0.690\cr\hline

    Feature$_{Social}$&0.616&0.678&0.703&0.729&0.742 &-&-&-&-&-\cr
    Feature$_{Forum}$&-&-&-&-&-& 0.491&0.543&0.563&0.581&0.595\cr\hline
    Graph$_{Factor}$&0.703&0.765&0.782&0.796&0.806& 0.643&0.679&0.692&0.702&0.710\cr
    Graph$_{GCN}$&0.765&0.802&0.824&0.850&0.865& 0.696&0.713&0.731&0.747&0.776\cr
    SplitGNN&0.742&0.798&0.820&0.847&0.860& 0.683&0.704&0.722&0.752&0.769\cr
    GFD&0.770&0.805&0.827&0.861&0.868& 0.690&0.726&0.733&0.757&0.773\cr\hline
    
    Latent$_{SVD}$&0.648&0.685&0.742&0.769&0.783& 0.594&0.602&0.645&0.664&0.671\cr
    Latent$_{GAUC}$&0.714&0.735&0.775&0.798&0.819& 0.641&0.658&0.697&0.710&0.735\cr
    Latent$_{n-aware}$&0.737&0.750&0.785&0.804&0.821& 0.652&0.661&0.698&0.736&0.740\cr\hline
    \end{tabular}}
\end{table*}

In this section, we comprehensively evaluate four aspects of our LFM models: effectiveness, applicability, stability of dealing with class imbalance, and tolerance in the face of the incomplete network.

\subsection{Datasets}
We start by introducing two real-world datasets used in this work. As we mentioned before, LFM models are capable of filtering various spamming activities resorting to the unified signed networks transformed from original networks of online platforms. To evaluate this generality, we select two disparate datasets from two types of Web applications --- the microblogging service and Web forum. These datasets are contaminated by different malicious activities, \textit{i.e.}, fake following relationships and deceptive opinions. If our methods (MRLE and SPR) achieve considerable effectiveness in dealing with these two categories of spamming campaigns, the usability and generality of signed latent factors in dealing with spam problems can be validated. 

The first dataset \textbf{\texttt{Social}} is collected by \cite{liu2016pay}, focusing on combating the fake follower problem \cite{almaatouq2014twitter} on social networking sites. The authors buy about 3K fake followers from the underground market, which are directly treated as a ground-truth spammer dataset. They then use empirical evidences to collect a non-spam user set. In order to obtain a relatively complete network, all the neighbours (followers and followees) of two kinds of user sets (spammers and non-spam users) are crawled. On microblogging platforms, each user has two social identities, \textit{i.e.}, follower and followee. To construct the unified signed network, we regard the followee identity of an account as the target node. It means that targets (followees) are linked to social user (follower) nodes by following relationships. Our objective is to detect fake following links. As following activities (links) are not directly labeled in \textbf{\texttt{Social}}, we regard the links between labeled spammers (fake followers) and suspicious followees (followed by more than one fake followers) as spam ones. 

The second dataset \textbf{\texttt{Forum}} is crawled for studying the opinion spam problem in Web forum \cite{chen2015opinion}, \textit{i.e.}, filtering deceptive contents that are used to promote or demote malicious products/services. A real case: Samsung probed over ‘fake web reviews’ in a popular Web forum is revealed. The authors gather all the certain deceptive posts on the revealed list and non-spam posts in the same time window. The dataset includes 10K first posts (5\% spams), 148K replies, and 60K user accounts (0.52\% spammers). The first post in the Web forum is like a topic/question, and other users can reply to it with user-generated contents. In this work, we focus on filtering deceptive replies (posted by labeled spammers) that link responders (users) to the first posts (targets). 

\subsection{Compared Methods} 
In Section 2.1, five types of application strategies of signed latent factors are described, which can be generally summarized into two types: 
\begin{itemize}[leftmargin=*]
\item We directly use signed latent factors to construct feature sets that are fed into a popular classification model (logistic regression), \textit{i.e.}, the application operators of \textit{Avg}, \textit{Con}, and \textit{Sub};
\item Referring to recommendation methods based on matrix factorization, we rank the inner product results of spam latent factors ($IP^-$ operator) and select top-ranked activities as spam ones. If the $IP^+$ operator is used, the low ranked links are regarded as spamming ones. 
\end{itemize}

To show that LFM models outperform existing mainstream methods, multiple baselines are used to compare and analyze in experiments: 
\begin{itemize}[leftmargin=*] 
\item pruned MRLE model --- MRLE$-non$ (\textit{i.e.}, ignoring non-linked relations), and improved MRLE model --- MRLE$+au$ that regards high ranking interactions identified by SPR as auxiliary known spamming links; 
\item graph propagation-based approaches applied to both \textbf{\texttt{Social}} and \textbf{\texttt{Forum}}, HITS-like detection method --- Graph$_{Hits}$ proposed by \cite{liu2016pay} and bipartite graph propagation model with polished edge weight --- Graph$_{Bi}$ \cite{zhang2015catch}; 
\item feature-based models, utilizing powerful logistic regression model to classify spamming activities based on common used features for fake follower detection \cite{aggarwal2015they,liu2016pay} (Feature$_{Social}$) and specific features extracted to combat opinion spam in Web forum \cite{chen2015opinion,jindal2010finding} (Feature$_{Forum}$);
\item graph and feature combined baseline --- Graph$_{Factor}$ is developed based on the probabilistic factor graph model \cite{liu2017detecting}, integrating correlation factors and attribute factors. Since recently developed Graph Convolutional Networks (GCNs) \cite{kipf2016semi, defferrard2016convolutional} can combine both graph structures and node features for semi-supervised learning, extensive studies resort to GCNs in spam detection tasks \cite{agarwal2022modeling, burkholder2021certification, wu2020graph}. We select the state-of-the-art GCN-based opinion spam detection model  \cite{li2019spam} as another baseline (Graph$_{GCN}$). 
In addition, we have incorporated two of the recent proposed baselines, SplitGNN \cite{wu2023splitgnn} and GFD \cite{zhuo2024partitioning}, which utilize a spectral graph neural network and partitioning message passing based GNN, respectively, to enhance spam detection.
\end{itemize}

Three state-of-the-art latent-factor methods are selected as baselines to indicate that addressing the spam detection problem using latent factors is a unique and innovative work, and our LFM models have distinct advantages on combating spam over these baselines. 
\begin{itemize}[leftmargin=*] 
\item Latent$_{SVD}$ (unsigned factors, $d = 30$) is conducted by the widely adopted Singular Value Decomposition (SVD) method \cite{abdi2007singular} on the original user-target matrix. It means that the sign of each entry is not considered, and latent factors of all the instances are mined based on the interaction network structure;
\item Latent$_{GAUC}$ (unsigned factors, $d = 30$) quantify the ranking performance in signed networks by focusing on the head (spam) and tail (normal) of a ranking list \cite{song2015recommending}. A low-rank model is employed to obtain ranking scores;
\item A negative-aware MF recommendation approach\cite{lin2019negative} explicitly addresses the uncertainty of unseen user-item associations, \textit{i.e.}, quantifying the degree of uncertainty for unseen associations by leveraging user preference similarity.We develop Latent$_{n-aware}$ (signed factors) by separately utilizing spam and normal associations to calculate spam- and normal-oriented user preference similarities. 
\end{itemize}

We make the latent factors of latent-factor methods and the concatenated embeddings learned from the GCN baseline share the same dimension (\textit{i.e.}, $d^+ + d^-$) with signed latent factors of SPR and MRLE. 

\begin{figure}
\centering
    \subfigure[\texttt{Social} (30\%)]{
    \includegraphics[width=0.48\linewidth]{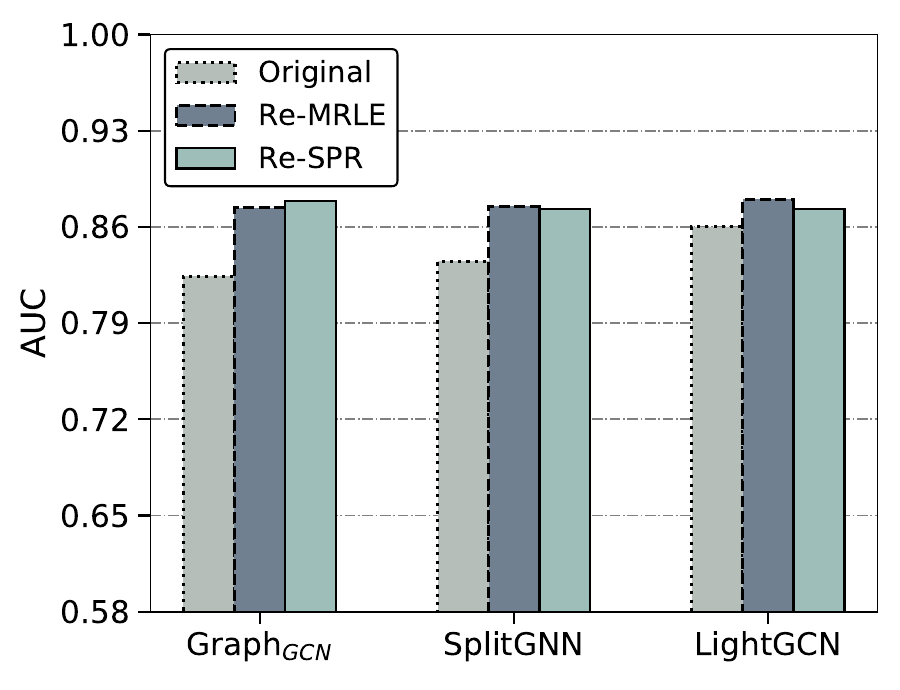}}
    \subfigure[\texttt{Forum} (30\%)]{
    \includegraphics[width=0.48\linewidth]{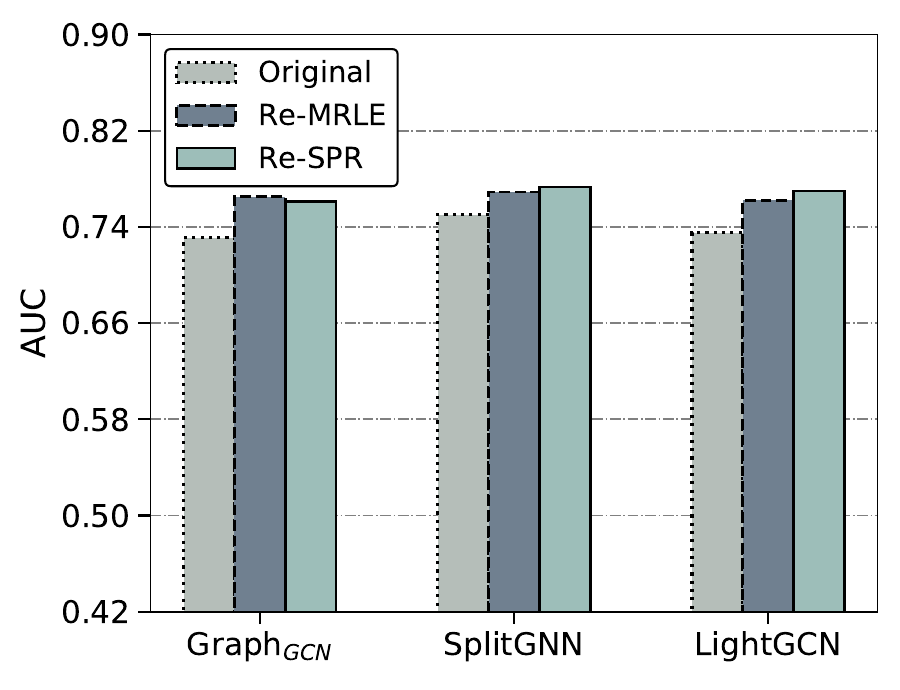}}
\caption{Affects of LFM optimization for neural models.}
\label{fig:applica}
\end{figure}

\subsection{Effectiveness}
Because we use a new perspective to solve the spam problem, it is important to detail how to set the training and test datasets. For example, 20\% labeled data in an experiment means that twenty percent of labeled spamming activities in the whole dataset are randomly selected, \textit{i.e.}, making their labels known to SPR and MRLE models in the learning process. To keep class balance, the same number of normal instances as the amount of selected 20\% spamming activities are selected. As for the remaining instances (not in the training dataset), they are label unknown for the SPR model (\textit{i.e.}, with a symbol of $?$ in the signed matrix). And MRLE model knows that these links are existed in the signed network but does not divide them into any of the three types of relational pairs. To fairly compare the mentioned methods above, they all utilize the same size of prior knowledge (\textit{i.e.}, the same training dataset). 

Table 2 reports the AUC performances of all the compared methods over two datasets. The useful and important observations derived from this table can be summarized into following aspects: 
\begin{itemize}[leftmargin=*]
\item Two types of signed latent factors mined from MRLE and SPR both surely provide effective potential indications for detecting spam campaigns, because five operators for applying LFM all play relatively positive detection roles. Among designed operators, the $Con$ obtains the best detection performance on both datasets with different sizes of varying labeled data compared to the other operators. In the following contents, SPR and MRLE represent the LFM models using the $Con$ operator;
\item Without using null relationships (MRLE$-non$), the detection performances will be negatively affected, which indicates the indispensable effects of seldom-mentioned $non$ relations. It's important to note that the improved MRLE model (MRLE$+au$) achieves the best performance in most cases by taking full advantages of non-linked relations and auxiliary ranking information of SPR;
\item Bipartite propagation (Graph$_{Bi}$) and HITS-like (Graph$_{HITS}$) methods achieve better performances than the other feature-based compared approaches, especially in the case of using the small size of prior knowledge (\textit{e.g.}, 20\%). This is mainly because propagation model makes use of the inherent connections between spammers and targets \cite{zhang2015catch, rayana2015collective} on the whole graph; 
\item The proposed SPR and MRLE show obvious superiority over the competitive GCN-based detection model (\textit{i.e.}, Graph$_{GCN}$, SplitGNN, and GFD) on \textbf{\texttt{Social}} and \textbf{\texttt{Forum}}, especially when using limited labeled datasets ($<40\%$). These results suggest that signed latent factors mining is conducive to the detection of spamming activities. Compared to these neural network based models that filters spam by propagating messages on known graph, LFM models fit naturally into the spam detection tasks by exploring critical information (null relationships and label unknown links). 
\item Even if a small proportion of labeled data is used as prior knowledge, our LFM models achieve relatively high AUC values compared with other methods. It demonstrates that the signed latent factors have the ability to deal with the insufficient labeled dataset by mining more attributes that are hidden. Besides, with the increase of training data, the performances of LFM models improve smoothly, which means that our methods are relatively stable;
\item The SPR model that considers label unknown activities performs better than MRLE when a few labeled datasets (20\% or 30\%) are used in \textbf{\texttt{Social}} and \textbf{\texttt{Forum}}. Two factors contribute to this result, 1) label unknown instances bring additional evidence for comparing ranking relations which are especially useful when the labeled data is limited, and 2) SPR model makes full use of the labeled data through repeatedly comparing them with others; 
\item Three latent-factors baselines use the $con$ operator to represent links. The Latent$_{SVD}$ does not perform well, as SVD latent factors are mined based on the network structure ignoring the sign of each entry. Although Latent$_{n-aware}$ provides added information for the unseen interaction by quantifying the degree of uncertainty and Latent$_{GAUC}$ considers signed links, they both are not properly tailored to make full use of the special \textbf{label unknown} and \textbf{null} relations in spam detection tasks. 
\end{itemize}

We also perform experiments using the combined method --- MRLE-SPR, which concatenate all signed latent factors mined from MRLE and SPR. The detection results of  MRLE-SPR are better than the individual models, increasing by 2.4\% and 3.1\% compared to MRLE and SPR, respectively, using \textbf{\texttt{Social}} (30\% labeled data, $n = 20$, $d = 30$, $\xi = 2$). It shows that the combined model MRLE-SPR learns from each other's strong points and provides more potentials.

\subsection{Applicability}

To validate the capability of our Latent Factor Mining (LFM) method to enhance existing neural network-based representation learning models in the spam detection domain, we selected three representative approaches for comparative analysis: (i) Graph$_{\text{GCN}}$ \cite{li2019spam}, a seminal work that utilizes Graph Convolutional Networks for opinion spam detection; (ii) SplitGNN \cite{wu2023splitgnn}, a recently proposed method employing spectral graph neural networks to capture appropriate signals for fraud detection; and (iii) LightGCN \cite{he2020lightgcn}, a competitive and simplified graph neural network that effectively leverages the structural information within graph networks to learn targeted representations. The results, illustrated in \Cref{fig:applica}, compare the original performance of these models on relevant datasets (original) with their performance after integrating our signed latent factors as trainable representations, subsequently optimized using our corresponding LFM method (Re-MRLE and Re-SPR). In this context, positive and negative latent factors separately learn from their corresponding interaction structures within signed networks.
The enhancements provided by our LFM approach are evident from \Cref{fig:applica}, demonstrating that both MRLE and SPR significantly improve the original neural network-based methods for spam detection. This improvement confirms that LFM can serve effectively as an optimization criterion. By leveraging neural network architectures to deeply model complex structural information, our approach further captures the distinct characteristics of normal and fraudulent entities, thereby elevating the detection capabilities of the models. This compelling evidence underscores the applicability of LFM, validating its ability to refine and advance the field of spam detection through strategic optimization of signed latent representations.

\subsection{Properties}

As is introduced in the formulations of MRLE and SPR, several parameters are used, such as $n$ (size of unconnected pairs for MRLE method), $d^-$ and $d^+$ (dimensions of positive and negative latent factors), $p_0$ in activation function of MRLE, learning rate $\alpha$ of SPR, regularization parameters $\lambda_{\Theta^-}$ and $\lambda_{\Theta^+}$ of SPR, and $\xi$ using as the identification threshold for SPR. By multiple experiments and empirical evidences, we initialize $p_0 = 0.01$, $\alpha = 0.005$, and $\lambda_{\Theta^-} = \lambda_{\Theta^+} = 0.01$ for two experimental datasets.      

\begin{figure}
\centering
    \subfigure[Size of non-linked relations]{
    \includegraphics[width=0.484\linewidth]{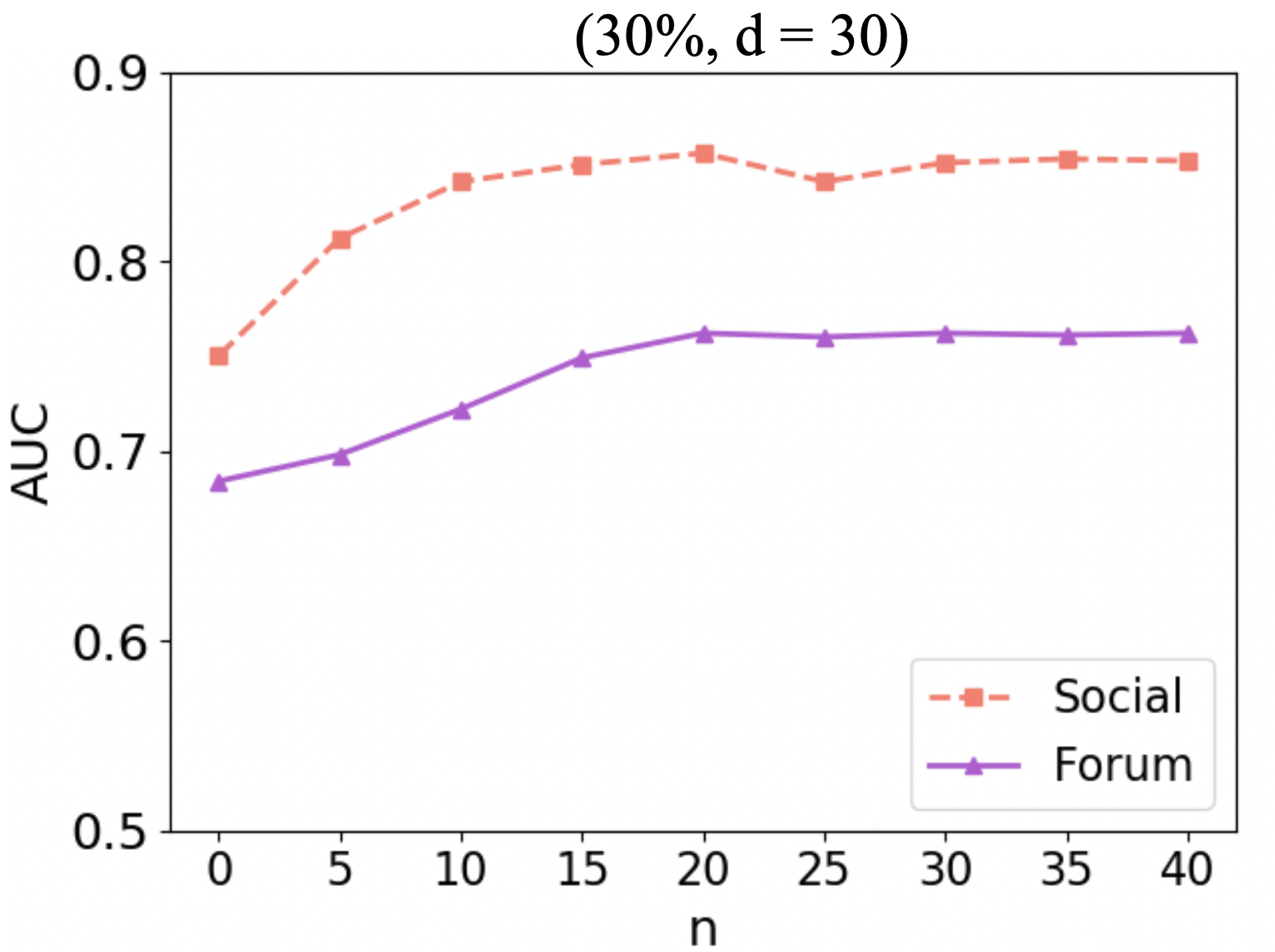}}
    \subfigure[Dimensions]{
    \includegraphics[width=0.484\linewidth]{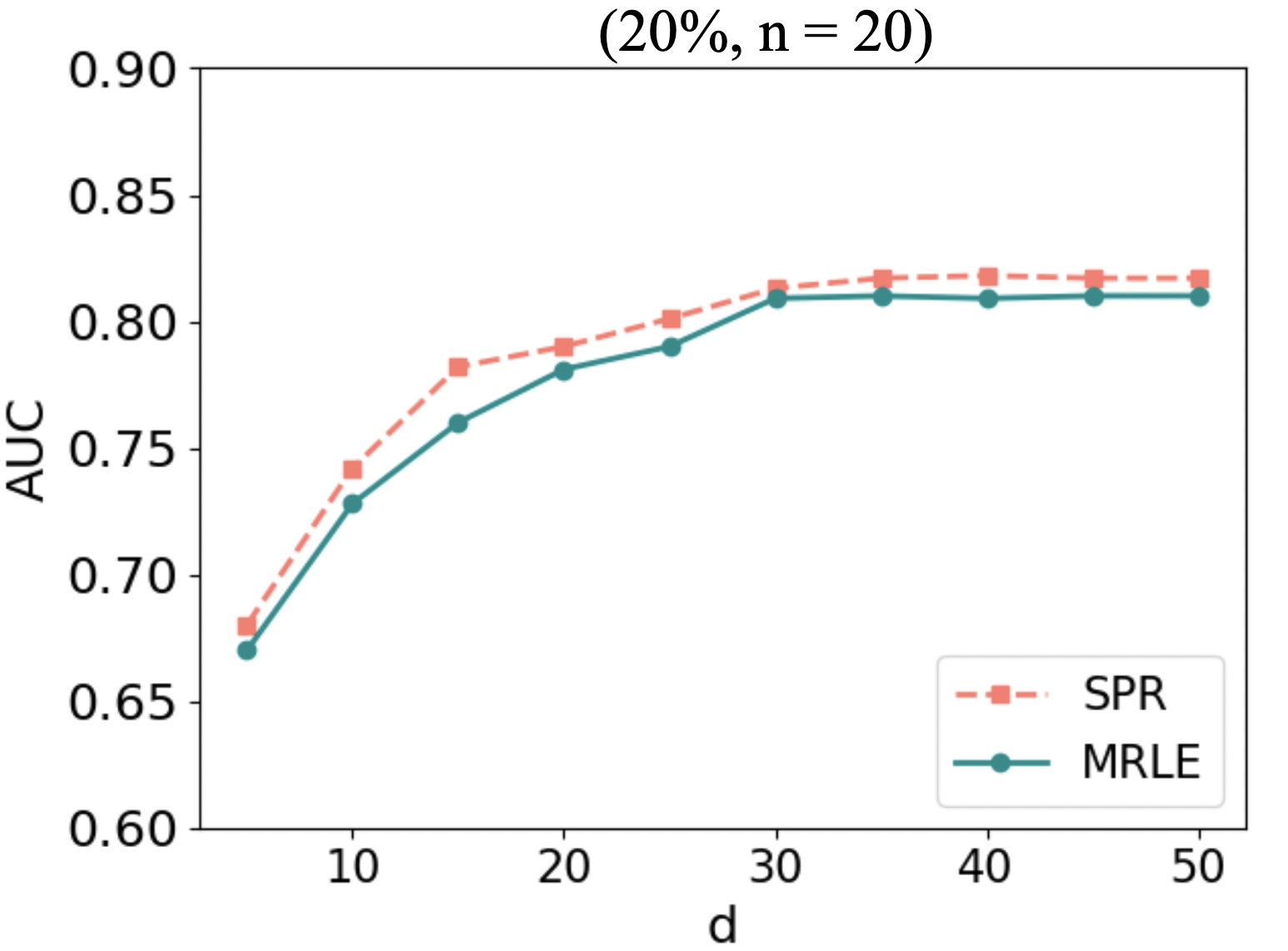}}
\caption{The impact of parameters $n$ and $d$.}
\label{fig:4}
\end{figure}

To gain financial benefit, spammers mainly perform activities that are linked to malicious targets. It means that the \textbf{disguised-activities} which are used by spammers to cheat detection models through building connections to normal targets are scattered with a relatively small number \cite{liu2020recommending}. Therefore, it is not feasible to assign a big value to the identification threshold $\xi$ (mentioned in Section 4.1). Through multiple experiments, we find that setting the value of $\xi$ smaller than 3 is advisable. If the value of $\xi$ is too large, the improvement of LFM models will be impeded. To save space, the influence of parameter $\xi$ for detection performance is not demonstrated, but we find that our experiments on two datasets achieve the best performance when $\xi=2$ (shown in Table 2). On the other hand, the sizes of non-linked pairs ($n$) and dimensions ($d$) of latent factors have important indicative influence for detecting spamming activities. We therefore use \Cref{fig:4} to investigate the effects of $n$ and $d$. In Table 2, the results of LFM models are obtained in the case of setting $n= 20$ and $d^- = d^+ = 30$. As we can separately calculate normal and spam latent factors, it is feasible to assign $d^-$ and $d^+$ with different values. However, due to the insignificant effects of setting different dimensions for positive and negative latent factors in our experiments, we assign the same value to $d^-$ and $d^+$ (\textit{i.e.}, $d$ ). As \Cref{fig:4} shows, the detection performances are sensitive to the size of $n$ and $d$ when the size values are relatively small, but corresponding influences on detection performances are gradually reduced with the increase of the values of $n$ and $d$. 

\begin{figure}
\centering
    \subfigure[Fake followers]{
    \includegraphics[width=0.484\linewidth]{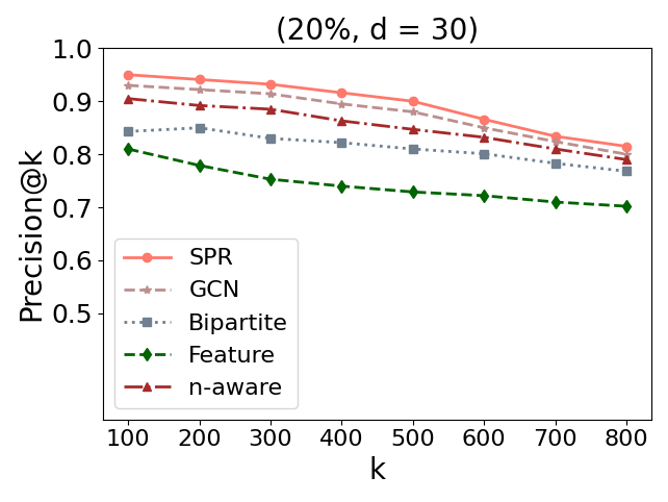}}
    \subfigure[Opinion spam]{
    \includegraphics[width=0.484\linewidth]{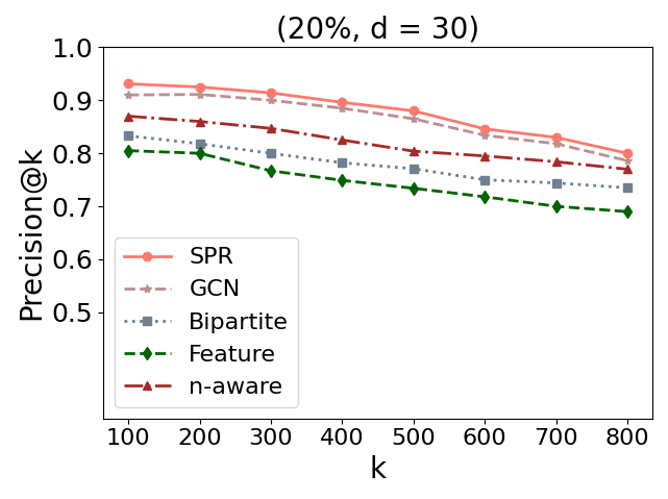}}
\caption{Precision@k of compared methods.}
\label{fig:5}
\end{figure}

To further evaluate and verify the effectiveness of our models, another common metric precision@k for spam detection tasks is used and the related results are shown in \Cref{fig:5}. The major findings learned from \Cref{fig:5} are: (\textbf{\romannumeral1}) LFM models outperform other competitive methods on two datasets \textit{w.r.t.} precision@k metric; (\textbf{\romannumeral2}) More than 90\% of activities in the top 500 ranking results detected by our models are deceptive. We therefore can say that solely relying on the signed latent factors to filter high-ranking instances as spamming results achieves very low error. This type of precision is quite important for spam detection tasks, because it means that normal instances are unlikely to be identified as spam ones.

\subsection{Dealing with Long-standing Challenges}

\begin{table*}[tp]
\centering
  \fontsize{7.5}{9.9}\selectfont
  \caption{F-measure performances in dealing with class imbalance ( \% labeled data, $n = 20$, $d = 30$).}
    \begin{tabular}{cccccccc|ccccccc}
    \hline
    \multirow{2}{*}{\%}&\multicolumn{7}{c|}{\textbf{\texttt{Social}}} & \multicolumn{7}{c}{\textbf{\texttt{Forum}}}\cr\cline
    {2-15}&MRLE&SPR&Bigraph&HITS&Feature&Factor& n-aware &MRLE&SPR& Bigraph & HITS&Feature& Factor & n-aware \cr\hline
    \ 20 \   
        &0.620&0.629&0.492&0.440&0.334 &0.529 &0.565
        &0.533&0.535&0.434&0.447&0.301 &0.468 &0.507 \cr\hline
    25  &0.635&0.640&0.535&0.485&0.398 &0.551 &0.570
        &0.550&0.543&0.501&0.472&0.332 &0.503 &0.516
        \cr\hline
    30  &0.649&0.646&0.576&0.527&0.417 &0.580 &0.594
        &0.582&0.579&0.548&0.493&0.367 &0.534 &0.556
        \cr\hline
    40   &0.664&0.658&0.593&0.549&0.437 &0.602 &0.615
        &0.597&0.591&0.570&0.531&0.378 &0.574 &0.589
        \cr\hline
    50  &0.670&0.665&0.601&0.564&0.460 &0.615 &0.637  
        &0.608&0.602&0.584&0.556&0.403 &0.590 &0.601
        \cr\hline
    \end{tabular}
\end{table*}

\begin{table*}[tp]
\centering
  \fontsize{7.5}{9.9}\selectfont
  \caption{F-measure in face of varying degrees of network incompleteness (40\% labeled data, $n = 20$, $d = 30$).}
     \begin{tabular}{cccccccc|ccccccc}
    \hline
    \multirow{2}{*}{$Degree$}&\multicolumn{7}{c|}{\textbf{\texttt{Social}}} & \multicolumn{7}{c}{\textbf{\texttt{Forum}}}\cr\cline
    {2-15}&MRLE &SPR&Bigraph&HITS&Feature&Factor& n-aware &MRLE&SPR& Bigraph & HITS&Feature& Factor & n-aware \cr\hline
    0\% &0.664&0.658&0.593&0.549&0.437 &0.602&0.615 
        &0.597&0.591&0.570&0.531&0.378 &0.574&0.589
    \cr\hline
    3\% &0.659&0.635&0.574&0.541&0.431 &0.585&0.609
        &0.590&0.586&0.561&0.516&0.370 &0.567&0.570
    \cr\hline
    5\%&0.647&0.630&0.556&0.527&0.422 &0.569&0.587 
       &0.584&0.582&0.537&0.508&0.361 &0.541&0.562
    \cr\hline
    7\%&0.641&0.622&0.536&0.503&0.414 &0.550&0.564
       &0.570&0.563&0.509&0.482&0.355 &0.518&0.537
    \cr\hline
    10\%&0.627&0.608&0.502&0.462&0.393 &0.527&0.531 
        &0.548&0.540&0.476&0.460&0.340 &0.483&0.505
    \cr\hline
    \end{tabular}
\end{table*}

Three long-standing challenges --- the lack of training labels, class imbalance, and network incompleteness are mentioned in the Introduction section. By varying labeled data used in the training process, the experimental results in Table 2 have indicated that LFM models perform well in coping with limited spam instances. When only 20\% labeled instances in \textbf{\texttt{Social}} are selected as training dataset, the MRLE$_{Con}$ still achieves 0.809 AUC. To evaluate our models' ability in confronting the other two challenges, specialized experiments are designed. 

It is important to notice that in the spam detection area, the widely used evaluation method of area under the ROC curve (AUC) is less sensitive to imbalanced data compared to the other metrics for spam ranking tasks. To comprehensively evaluate our models in the face of class imbalance, we further apply the F-measure that synthesizes precision and recall measures, providing more precise and rounded illustrations of a detection method. 

In this work, both the training and test class imbalances are considered by employing specific experimental setups: (\romannumeral1) To simulate different degrees of training class imbalance, we fix the number of normal activities in the training dataset which is the same as the number of 50\% spam instances. In this way, reducing the percentage of selected spam labeled data (less than 50\%) will cause training class imbalance, \textit{i.e.}, the fewer amount (\%) of spam instances are selected the more severe of class imbalance; (\romannumeral2) As for test class imbalance, the ratio of spam to normal activities is fixed to 1:10. For example, in an experiment, 80\% of spamming activities (2K) are not selected as training data but are used for testing. We then randomly choose 20K instances that are not labeled spamming activities and are not used in the training process as normal ones for evaluation. To save space, only the results of the negative-aware method Latent$_{n-aware}$ are listed in the following tables, which achieves the best performance among three latent-factors baselines. 

Table 3 shows detection results of all compared methods in dealing with different levels of class imbalance. The indications from Table 3 are summarised as: 
\begin{itemize}[leftmargin=*]
\item Even if the F-measure is used for evaluation, LFM models still outperform other baselines in the face of different degrees of class imbalance; 
\item Compared with structure-based approaches, feature-based methods are not strong enough to deal with severe class imbalance; 
\item By analyzing the changing trend of each method’s performance using different amounts of seeds, the F-measure performances of our models are more stable than the other ones. 
\end{itemize}

We use Table 4 to report compared methods' performances of tolerating varying degrees of network incompleteness. To evaluate this, we randomly remove different proportions (\%) of links (\textit{i.e.}, activities) from the original networks. We can find that compared to the results on the whole complete dataset (\textit{i.e.}, 0\%), all the experiments on more ``incomplete'' networks obtain inferior performances, which means that the incompleteness of graph poses an obstacle to detection models. However, the effectiveness drops of LFM methods are lower than other models when increasing the degree of incompleteness, \textit{i.e.}, our models achieve better tolerance in face of the incomplete graph.

\section{RELATED WORK}
Three lines of studies are closely related to this work: spam detection, signed social networks, and pairwise learning.
\subsection{Spam Detection}
The most commonly used spam detection method is to utilize machine learning to train a classifier to distinguish spam instances from normal ones. The differences among the classifiers mainly lie in the features used to represent the data \cite{liu2017detecting}. It’s generally recognized that there are clues for spotting spam in user-generated contents. Kim \textsl{et al.} \cite{kim2015deep} deploy a deep framework to identify linguistic features for detecting opinion spam. The other content-based features (\textit{e.g.}, semantic representation and syntactic structure) are also exploited to combat deceptive reviews \cite{kim2015deep, sandulescu2015detecting} or promotion campaigns \cite{liu2017detecting, liu2016detecting}.  Besides, employing user rating patterns \cite{lim2010detecting, kumar2018rev2} and clicking behaviors \cite{shah2017flock, su2018detecting} for spam detection is also quite common. 
To combat well-disguised crowdturfing activities, many existing works attempt to mine collusive signals, which can be implemented from two aspects: (\romannumeral1) capturing synchronized behaviors \cite{li2016modeling, jiang2014catchsync}; (\romannumeral2) grouping spammers or spamming activities \cite{xu2013uncovering, fayazi2015uncovering}.
By utilizing inherent relationships between spammers and malicious targets, structure-based approaches achieve relatively good performances in detecting spam, which are mainly divided into two types: probabilistic graphical models \cite{su2018detecting, lu2013simultaneously} and label propagation methods \cite{zhang2015catch, liu2016pay}. 
In this work, the signed latent factors are mined based on relational ties, which means that our method can be seen as a type of structure-based approach. However, the signed latent factor models are capable of uncovering plentiful potential factors from various online networks, which are fed into a classification algorithm to achieve spam detection. On the other hand, MRLE and SPR utilize the null relationships and label unknown links, respectively. And the process of comparing pairs in SPR is relatively random (different from other structure-based methods whose inference process is guided by edges). We can construct enough comparing ranking pairs based on existing entries ($+$, $-$ and $?$) to ensure acceptable detection performance.

The interpretability of spam detection models is crucial for understanding the underlying mechanics of decision-making processes, especially in applications where trust and accountability are paramount. Feature-based methods excel in interpretability as they rely on explicit attributes derived from the data. These attributes can be directly linked to the input features, making it easier for practitioners to understand why certain decisions are made \cite{yuan2016interpretable, 10574870}. 
In contrast, LFM leverages latent factors to capture underlying patterns in the data that are not immediately observable. These latent factors, while powerful in handling complex data, typically do not offer a clear one-to-one correspondence with interpretable features. 
However, the interpretability of latent factor models can be enhanced through techniques such as factor analysis \cite{alvarez2024comprehensive}, mapping latent dimensions back to known features post hoc \cite{stites2021sage}, and developing hybrid models that integrate both interpretable features and latent representations. Enhancing the interpretability of LFM will be a focus of our future work \cite{dias2021hybrid}. 

\subsection{Signed Social Networks}
Many real-world relations can be represented by signed networks with positive and negative links, \textit{e.g.}, the positive/negative election of Wikipedia admins \cite{cabunducan2011voting} and trust/distrust \cite{leskovec2010predicting} in Epinions. Comparing unsigned networks, two unique properties are presented in signed networks distinct properties and collective properties \cite{tang2016survey}, which involve more information and facilitate mining tasks (\textit{e.g.}, node classification, link prediction recommendation, etc.). Although there are many similar tasks for mining signed and unsigned networks, the task of sign prediction (inferring the signs of existing links) is unique to signed networks. 
This work is among the first to treat spam contaminated relational datasets as signed social networks with spam and normal links between users and targets. Based on that treatment, we design the spam detection-oriented signed latent factor models.

\subsection{Pairwise Learning}
Rendle \textsl{et al.} \cite{rendle2009bpr} propose the pairwise learning algorithm --- BPR which is optimized based on the relative preference of a specific user over pairs of items. BPR is a highly competitive approach in the item recommendation area and has been utilized in many papers as the state-of-the-art baseline up until recently \cite{he2017neural, liu2024learning}. In this paper, our SPR model is inspired by the core thought of BPR method in the recommendation area. However, there are significant differences between SPR and BPR: (\romannumeral1) SPR makes full use of the existing relations between users and targets (including unlabeled and test instances in our dataset) and designs three types of entries according to the signed network; (\romannumeral2) Comparing pairs are constructed from two levels (spam and normal) based on user-oriented perspective by analyzing potential indications; (\romannumeral3) Each user and item acquire two types of latent factors, \textit{i.e.}, spam (negative) and normal (positive) potential representations. 

\section{CONCLUSIONS}
Detecting online spamming activities has been extensively studied. In this work, we present a new thought to tackle this popular and intractable problem by introducing the signed latent factors mining into the spam detection area. To provide more potentials and possibilities, two signed latent factor mining models from two quite different perspectives (pointwise and pairwise) are proposed, which are both effective and perform well in coping with the long-standing challenges in combating spam. We comprehensively explore how to represent activities (links) using signed user and target latent factors, by proposing and evaluating five types of application operators. In order to improve the generality, we design the unified network form to represent different kinds of contaminated relational data on various Web platforms. By performing experiments on two disparate datasets, the respectable usability and generality for spam detection of LFM models are validated. Besides, we believe that the LFM thought can also be extended to other spam detection tasks, \textit{e.g.}, fraud view problem in online video platforms (user-video network) \cite{shah2017flock} and deceptive review detection in online shopping sites (user-product matrix) \cite{jindal2008opinion}. In the future, we will evaluate our models on other tricky spam detection tasks and explore the interpretability within the context of latent factors based detection approach.

\ifCLASSOPTIONcaptionsoff
  \newpage
\fi



%
\bibliographystyle{IEEEtran}
\bibliography{reference_sample}

%

\begin{IEEEbiography}[{\includegraphics[width=1in,height=1.25in,clip,keepaspectratio]{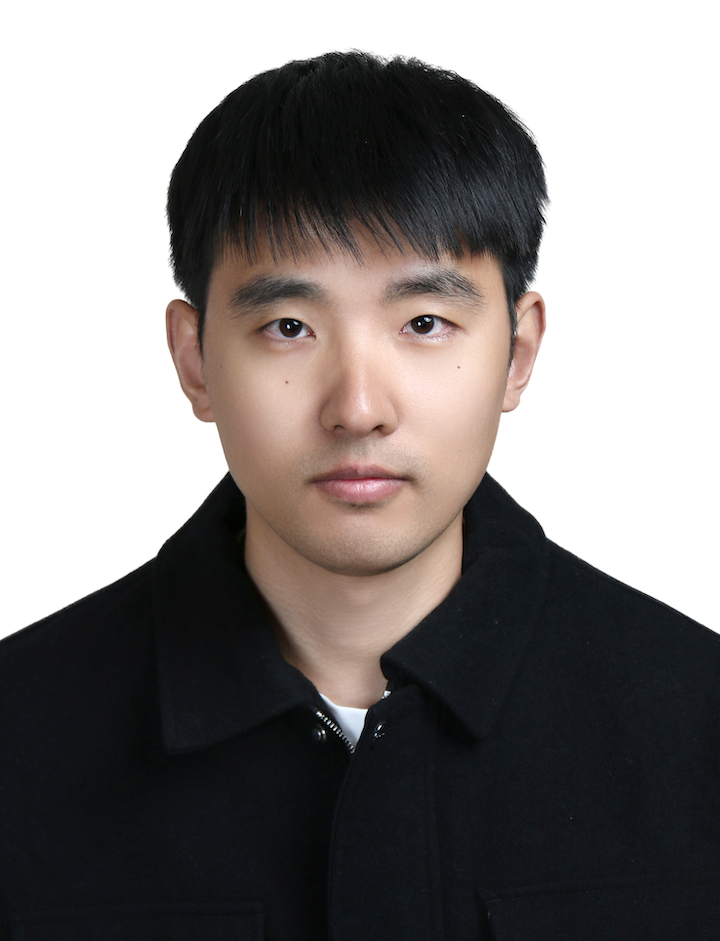}}]{Yuli Liu}
received the master degree in Computer Science and Technology from Tsinghua University, and the PhD degree at the College of Engineering, Computing and Cybernetics, in The Australian National University. He is currently working in Qinghai University and Quan Cheng Laboratory. His research interest includes data mining, information retrieval, spam detection.
\end{IEEEbiography}







\end{document}